\RequirePackage[2020-02-02]{latexrelease}

\documentclass[twocolumn,preprintnumbers]{revtex4}%
\usepackage{amssymb}
\usepackage{amsmath}
\usepackage{graphicx}
\usepackage{dcolumn}
\usepackage{bm}
\usepackage{xcolor}
\usepackage{ifthen}
\usepackage[normalem]{ulem}
\usepackage{amsfonts}%
\RequirePackage[2020-02-02]{latexrelease}
\UseRawInputEncoding
\providecommand{\U}[1]{\protect\rule{.1in}{.1in}}
\providecommand{\U}[1]{\protect\rule{.1in}{.1in}}
\def\showal{1}
\newcommand{\al}[1]{\ifthenelse{\showal=1}{\textcolor{orange}{[[#1]]}}{}}

\newcommand{\eb}[1]{\ifthenelse{\showal=1}{\textcolor{cyan}{[[#1]]}}{}}
\begin{document}
\title{Spontaneous disentanglement and thermalisation}
\author{Eyal Buks}
\email{eyal@ee.technion.ac.il}
\affiliation{Andrew and Erna Viterbi Department of Electrical Engineering, Technion, Haifa
32000, Israel}
\date{\today }

\begin{abstract}
The problem of quantum measurement can be partially resolved by incorporating
a process of spontaneous disentanglement into quantum dynamics. We propose a
modified master equation, which contains a nonlinear term giving rise to both
spontaneous disentanglement and thermalisation. We find that the added
nonlinear term enables limit cycle steady states, which are prohibited in
standard quantum mechanics. This finding suggests that an experimental
observation of such a limit cycle steady state can provide an important
evidence supporting the spontaneous disentanglement hypothesis.

\end{abstract}
\maketitle





\section{Introduction}

The problem of quantum measurement
\cite{Schrodinger_807,von_Neumann_Mathematical_Foundations} is considered as
one of the most important open questions in physics. This problem arguably
originates from a self-inconsistency in quantum theory
\cite{Penrose_4864,Leggett_939,Leggett_022001}. This long-standing problem has
motivated some proposals for nonlinear extensions to quantum theory
\cite{Weinberg_61,Doebner_3764,Gisin_5677,Kaplan_055002,Munoz_110503,Geller_2200156}%
. Moreover, processes giving rise to spontaneous collapse have been explored
\cite{Bassi_471,Pearle_857,Ghirardi_470,Bassi_257,Bennett_170502,Kowalski_1,Fernengel_385701,Kowalski_167955,Oppenheim_041040,Schrinski_133604}%
. For some cases, however, nonlinear quantum dynamics may give rise to
conflicts with well-established physical principles, such as causality
\cite{Bassi_055027,Jordan_022101,Polchinski_397,Helou_012021,Rembielinski_012027,Rembielinski_420}
and separability \cite{Hejlesen_thesis,Jordan_022101,Jordan_012010}. In
addition, some predictions of standard quantum mechanics (QM), which have been
experimentally confirmed to very high accuracy, are inconsistent with some of
the proposed extensions.

A modified Schr\"{o}dinger equation having a nonlinear term that gives rise to
suppression of entanglement (i.e. disentanglement) has been recently proposed
\cite{Buks_2300103}. This nonlinear extension partially resolves the
self-inconsistency associated with the measurement problem by making the
collapse postulate of QM redundant. The proposed modified Schr\"{o}dinger
equation can be constructed for any physical system whose Hilbert space has
finite dimensionality, and it does not violate norm conservation of the time
evolution. The nonlinear term added to the Schr\"{o}dinger equation has no
effect on product (i.e. disentangled) states. The spontaneous disentanglement
generated by the modified Schr\"{o}dinger equation gives rise to a process
similar to state vector collapse.

The nonlinear extension that was proposed in Ref. \cite{Buks_2300103} is
applicable only for bipartite systems, and only for pure states. To allow
incorporating spontaneous disentanglement for more general cases, we propose
here a modified master equation for the time evolution of the density operator
$\rho$. The modified master equation [see Eq. (\ref{MME}) below] contains a
nonlinear term that gives rise to spontaneous disentanglement. For a
multipartite system, disentanglement between any pair of subsystems can be
introduced by the added nonlinear term. Moreover, thermalisation can be
incorporated by an additional nonlinear term added to the master equation
(\ref{MME}).

In contrast to the modified master equation (\ref{MME}), which is nonlinear in
$\rho$, in standard QM the time evolution of $\rho$ is governed by the
Gorini-Kossakowski-Sudarshan-Lindblad (GKSL) master equation
\cite{Fernengel_385701,Lindblad_119,Manzano_025106}, which is linear in $\rho
$. This linear dependency excludes any nonlinear dynamics in the time
evolution of $\rho$ (see appendix B of Ref. \cite{Levi_053516}), and, in
particular, it excludes a limit cycle steady state for any quantum system
having a Hilbert space of finite dimensionality and time independent
Hamiltonian. On the other hand, as is demonstrated below, the modified master
equation (\ref{MME}) yields rich nonlinear dynamics. In particular, both a
Hopf bifurcation and a limit cycle steady state may occur. Note, however,
that, nonlinearity in $\rho$ does not necessarily imply that dynamical
instabilities are possible, as was demonstrated in Ref. \cite{Buks_052217}.

\section{Modified master equation}

Consider a modified Schr\"{o}dinger equation for the ket vector $\left\vert
\psi\right\rangle $ having the form \cite{Buks_2300170}%
\begin{equation}
\frac{\mathrm{d}}{\mathrm{d}t}\left\vert \psi\right\rangle =\left(
-i\hbar^{-1}\mathcal{H}-\Theta+\left\langle \psi\right\vert \Theta\left\vert
\psi\right\rangle \right)  \left\vert \psi\right\rangle \;, \label{MSE}%
\end{equation}
where $\hbar$ is the Planck's constant, $\mathcal{H}^{{}}=\mathcal{H}^{\dag}$
is the Hamiltonian, and the operator $\Theta^{{}}=\Theta^{\dag}$ is allowed to
depend on $\left\vert \psi\right\rangle $. Note that the norm conservation
condition $0=\left(  \mathrm{d}/\mathrm{d}t\right)  \left\langle \psi\right.
\left\vert \psi\right\rangle $ is satisfied by the modified Schr\"{o}dinger
equation (\ref{MSE}), provided that $\left\vert \psi\right\rangle $ is
normalized, i.e. $\left\langle \psi\right.  \left\vert \psi\right\rangle =1$.

The modified Schr\"{o}dinger equation (\ref{MSE}) for the ket vector
$\left\vert \psi\right\rangle $ yields a master equation for the pure state
density operator $\rho=\left\vert \psi\right\rangle \left\langle
\psi\right\vert $ given by \cite{Grimaudo_033835,Kowalski_167955}%
\begin{equation}
\frac{\mathrm{d}\rho}{\mathrm{d}t}=i\hbar^{-1}\left[  \rho,\mathcal{H}\right]
-\Theta\rho-\rho\Theta+2\left\langle \Theta\right\rangle \rho\;, \label{MME}%
\end{equation}
where $\left\langle \Theta\right\rangle =\left\langle \psi\right\vert
\Theta\left\vert \psi\right\rangle =\operatorname{Tr}\left(  \Theta
\rho\right)  $. Note that $\mathrm{d}\operatorname{Tr}\rho/\mathrm{d}t=0$
provided that $\operatorname{Tr}\rho=1$ (i.e. $\rho$ is normalized), and that
$\mathrm{d}\operatorname{Tr}\rho^{2}/\mathrm{d}t=0$, provided that $\rho
^{2}=\rho$ (i.e. $\rho$ represents a pure state) [see Eq. (\ref{MME})].

For the case $\mathcal{H}=0$, and for a fixed operator $\Theta$, the modified
master equation (\ref{MME}) yields an equation of motion for $\left\langle
\Theta\right\rangle $ given by%
\begin{equation}
\frac{\mathrm{d}\left\langle \Theta\right\rangle }{\mathrm{d}t}=-2\left\langle
\left(  \Theta-\left\langle \Theta\right\rangle \right)  ^{2}\right\rangle \;.
\label{d<Q>/dt}%
\end{equation}
The above result (\ref{d<Q>/dt}) implies for this case that the expectation
value $\left\langle \Theta\right\rangle $ monotonically decreases with time.
Hence, the nonlinear term in the modified master equation (\ref{MME}) can be
employed to suppress a given physical property, provided that $\left\langle
\Theta\right\rangle $ quantifies that property.

Here the operator $\Theta$ is assumed to be given by $\Theta=\gamma
_{\mathrm{H}}\mathcal{Q}^{\left(  \mathrm{H}\right)  }+\gamma_{\mathrm{D}%
}\mathcal{Q}^{\left(  \mathrm{D}\right)  }$, where both rates $\gamma
_{\mathrm{H}}$ and $\gamma_{\mathrm{D}}$ are positive, and both operators
$\mathcal{Q}^{\left(  \mathrm{H}\right)  }$ and $\mathcal{Q}^{\left(
\mathrm{D}\right)  }$\ are Hermitian. The first term $\gamma_{\mathrm{H}%
}\mathcal{Q}^{\left(  \mathrm{H}\right)  }$, which gives rise to
thermalisation \cite{Grabert_161,Ottinger_052119}, is discussed below in
section \ref{SecT}, whereas section \ref{SecDE} is devoted to the second term
$\gamma_{\mathrm{D}}\mathcal{Q}^{\left(  \mathrm{D}\right)  }$, which gives
rise to disentanglement.

\section{Thermalisation}

\label{SecT}

Consider the master equation (\ref{MME}) for the case where $\mathcal{H}^{{}}$
is time independent, $\gamma_{\mathrm{D}}=0$ (i.e. no disentanglement), and
$\mathcal{Q}^{\left(  \mathrm{H}\right)  }=\beta\mathcal{U}_{\mathrm{H}}$,
where $\mathcal{U}_{\mathrm{H}}=\mathcal{H}+\beta^{-1}\log\rho$ is the
Helmholtz free energy operator, $\beta=1/\left(  k_{\mathrm{B}}T\right)  $ is
the thermal energy inverse, $k_{\mathrm{B}}$ is the Boltzmann's constant, and
$T$ is the temperature. For this case, the thermal equilibrium density matrix
$\rho_{0}$, which is given by%
\begin{equation}
\rho_{0}=\frac{e^{-\beta\mathcal{H}}}{\operatorname{Tr}\left(  e^{-\beta
\mathcal{H}}\right)  }\;,
\end{equation}
is a steady state solution of the master equation (\ref{MME}), for which the
Helmholtz free energy $\left\langle \mathcal{U}_{\mathrm{H}}\right\rangle $ is
minimized \cite{Jaynes_579,Grabert_161,Ottinger_052119,Buks_052217}. The rate
$\gamma_{\mathrm{H}}$ represents the thermalisation inverse time. Note that
$\gamma_{\mathrm{H}}$ needs not be a constant.

\section{Disentanglement}

\label{SecDE}

Consider the case where $\gamma_{\mathrm{H}}=0$ (i.e. no thermalisation). As
can be seen from Eq. (\ref{d<Q>/dt}), disentanglement can be generated by the
term proportional to $\gamma_{\mathrm{D}}$\ in the Schr\"{o}dinger equation
(\ref{MSE}), and by the term proportional to $\gamma_{\mathrm{D}}$\ in the
master equation (\ref{MME}), provided that the operator $\mathcal{Q}^{\left(
\mathrm{D}\right)  }$ is chosen such that $\left\langle \mathcal{Q}^{\left(
\mathrm{D}\right)  }\right\rangle $ quantifies entanglement
\cite{Schlienz_4396,Peres_1413,Hill_5022,Wootters_1717,Coffman_052306,Vedral_2275,Eltschka_424005,Dur_062314,Coiteux_200401,Takou_011004,Elben_200501}%
. In Ref. \cite{Buks_2300103} the operator $\mathcal{Q}^{\left(
\mathrm{D}\right)  }$ was chosen to be equal to $\mathcal{Q}^{\left(
\mathrm{S}\right)  }$, where $\mathcal{Q}^{\left(  \mathrm{S}\right)  }$ is
constructed using the Schmidt decomposition. This allowed deriving a modified
Schr\"{o}dinger equation having the form given by Eq. (\ref{MSE}), which
contains a nonlinear term that gives rise to pure state bipartite
disentanglement. However, the Schmidt decomposition is inapplicable for both
mixed states and for multipartite systems. Here we employ an alternative
operator (henceforth denoted as $\mathcal{Q}^{\left(  \mathrm{D}\right)  }$),
which can be used to derive a modified master equation having the form given
by Eq. (\ref{MME}), and which is applicable for a general multipartite case
\cite{Buks_2300170}, and for a general mixed state.

Consider a multipartite system composed of three subsystems labeled as 'a',
'b' and 'c'. The Hilbert space of the system $H=H_{\mathrm{a}}\otimes
H_{\mathrm{b}}\otimes H_{\mathrm{c}}$ is a tensor product of subsystem Hilbert
spaces $H_{\mathrm{a}}$, $H_{\mathrm{b}}$ and $H_{\mathrm{c}}$. The
dimensionality of the Hilbert space $H_{\mathrm{L}}$ of subsystem $\mathrm{L}%
$, which is denoted by $d_{\mathrm{L}}$, where $\mathrm{L}\in\left\{
\mathrm{a},\mathrm{b},\mathrm{c}\right\}  $, is assumed to be finite. A
general observable of subsystem $\mathrm{L}$ can be expanded using the set of
generalized Gell-Mann matrices $\left\{  \lambda_{1}^{\left(  \mathrm{L}%
\right)  },\lambda_{2}^{\left(  \mathrm{L}\right)  },\cdots,\lambda
_{d_{\mathrm{L}}^{2}-1}^{\left(  \mathrm{L}\right)  }\right\}  $.

Entanglement between subsystems a and b can be characterized by the matrix
$D_{\mathrm{ab}}\equiv\rho_{\mathrm{ab}}-\rho_{\mathrm{a}}\otimes
\rho_{\mathrm{b}}$, where $\rho_{\mathrm{ab}}$ is the reduced density matrix
of the combined a and b subsystems, and $\rho_{\mathrm{a}}$ ($\rho
_{\mathrm{b}}$) is the reduced density matrix of subsystem a (b). The
following holds [see Eq. (\ref{D=}) of appendix \ref{AppHSF}]%
\begin{equation}
D_{\mathrm{ab}}=\sum_{a=1}^{d_{\mathrm{a}}^{2}-1}\sum_{b=1}^{d_{\mathrm{b}%
}^{2}-1}\frac{\left\langle \mathcal{C}\left(  \lambda_{a}^{\left(
\mathrm{a}\right)  },\lambda_{b}^{\left(  \mathrm{b}\right)  }\right)
\right\rangle \lambda_{a}^{\left(  \mathrm{a}\right)  }\otimes\lambda
_{b}^{\left(  \mathrm{b}\right)  }\otimes I_{\mathrm{c}}}{4}\;, \label{D_ab}%
\end{equation}
where for any given observable $O_{\mathrm{a}}^{{}}=O_{\mathrm{a}}^{\dag}$ of
subsystem a, and a given observable $O_{\mathrm{b}}^{{}}=O_{\mathrm{b}}^{\dag
}$ of subsystem b, the observable $\mathcal{C}\left(  O_{\mathrm{a}%
},O_{\mathrm{b}}\right)  $ is defined by%
\begin{equation}
\mathcal{C}\left(  O_{\mathrm{a}},O_{\mathrm{b}}\right)  =O_{\mathrm{a}%
}\otimes O_{\mathrm{b}}\otimes I_{\mathrm{c}}-\left\langle O_{\mathrm{a}%
}\otimes I_{\mathrm{b}}\otimes I_{\mathrm{c}}\right\rangle \left\langle
I_{\mathrm{a}}\otimes O_{\mathrm{b}}\otimes I_{\mathrm{c}}\right\rangle \;,
\label{C(O_a,O_b)}%
\end{equation}
where $I_{\mathrm{L}}$ is the $d_{\mathrm{L}}\times d_{\mathrm{L}}$ identity
matrix, and where $\mathrm{L}\in\left\{  \mathrm{a},\mathrm{b},\mathrm{c}%
\right\}  $.

The above result (\ref{D_ab}) suggests that entanglement between subsystems a
and b can be quantified by the nonnegative variable $\tau_{\mathrm{ab}}$,
which is given by $\tau_{\mathrm{ab}}=\left\langle \mathcal{Q}_{\mathrm{ab}%
}^{\left(  \mathrm{D}\right)  }\right\rangle $, where the operator
$\mathcal{Q}_{\mathrm{ab}}^{\left(  \mathrm{D}\right)  }$ is given by%
\begin{equation}
\mathcal{Q}_{\mathrm{ab}}^{\left(  \mathrm{D}\right)  }=\eta_{\mathrm{ab}%
}\operatorname{Tr}\left(  C^{\mathrm{T}}\left\langle C\right\rangle \right)
\;, \label{Q_12 Tr}%
\end{equation}
and where $\eta_{\mathrm{ab}}$ is a positive constant. The $\left(
a,b\right)  $ entry of the $\left(  d_{\mathrm{a}}^{2}-1\right)  \times\left(
d_{\mathrm{b}}^{2}-1\right)  $ matrix $C$ is the observable $\mathcal{C}%
\left(  \lambda_{a}^{\left(  \mathrm{a}\right)  },\lambda_{b}^{\left(
\mathrm{b}\right)  }\right)  $, and the $\left(  a,b\right)  $ entry of the
$\left(  d_{\mathrm{a}}^{2}-1\right)  \times\left(  d_{\mathrm{b}}%
^{2}-1\right)  $ matrix $\left\langle C\right\rangle $ is its expectation
value $\left\langle \mathcal{C}\left(  \lambda_{a}^{\left(  \mathrm{a}\right)
},\lambda_{b}^{\left(  \mathrm{b}\right)  }\right)  \right\rangle $, and thus
$\tau_{\mathrm{ab}}$ can be expressed as [compare to Eq. (31) of Ref.
\cite{Schlienz_4396}]%
\begin{equation}
\tau_{\mathrm{ab}}=\eta_{\mathrm{ab}}\sum_{a=1}^{d_{\mathrm{a}}^{2}-1}%
\sum_{b=1}^{d_{\mathrm{b}}^{2}-1}\left\langle \mathcal{C}\left(  \lambda
_{a}^{\left(  \mathrm{a}\right)  },\lambda_{b}^{\left(  \mathrm{b}\right)
}\right)  \right\rangle ^{2}\;. \label{tau_12}%
\end{equation}
In a similar way, the entanglement between subsystems $\mathrm{b}$ and
$\mathrm{c}$, which is denoted by $\tau_{\mathrm{bc}}$, and the entanglement
between subsystems $\mathrm{c}$ and $\mathrm{a}$, which is denoted by
$\tau_{\mathrm{ca}}$, can be defined. Deterministic disentanglement between
subsystems $\mathrm{L}^{\prime}$ and $\mathrm{L}^{\prime\prime}$ can be
generated by the modified master equation (\ref{MME}), provided that the
operator $\mathcal{Q}^{\left(  \mathrm{D}\right)  }$ in Eq. (\ref{MME}) is
replaced by the operator $\mathcal{Q}_{\mathrm{L}^{\prime},\mathrm{L}%
^{\prime\prime}}^{\left(  \mathrm{D}\right)  }$.

The entanglement variable $\tau_{\mathrm{ab}}$ is invariant under any single
subsystem unitary transformation \cite{Schlienz_4396}. Under such a
transformation, the matrix $C$ is transformed according to $C\rightarrow
C^{\prime}=T_{\mathrm{a}}^{{}}CT_{\mathrm{b}}^{\mathrm{T}}$. A completeness
relation, which is satisfied by the generalized Gell-Mann matrices [see Eq.
(8.177) or Ref. \cite{Buks_QMLN}], can be used to show that both the $\left(
d_{\mathrm{a}}^{2}-1\right)  \times\left(  d_{\mathrm{a}}^{2}-1\right)  $
matrix $T_{\mathrm{a}}^{{}}$ and the $\left(  d_{\mathrm{b}}^{2}-1\right)
\times\left(  d_{\mathrm{b}}^{2}-1\right)  $ matrix $T_{\mathrm{b}}^{{}}$ are
orthonormal, i.e. $T_{\mathrm{a}}^{\mathrm{T}}T_{\mathrm{a}}^{{}%
}=T_{\mathrm{a}}^{{}}T_{\mathrm{a}}^{\mathrm{T}}=1$ and $T_{\mathrm{b}%
}^{\mathrm{T}}T_{\mathrm{b}}^{{}}=T_{\mathrm{b}}^{{}}T_{\mathrm{b}%
}^{\mathrm{T}}=1$, and thus, as can be seen from Eq. (\ref{Q_12 Tr}),
$\tau_{\mathrm{ab}}$ is invariant [see Eq. (8.881) of Ref. \cite{Buks_QMLN}].
The invariance of $\tau_{\mathrm{bc}}$ and\ $\tau_{\mathrm{ca}}$\ can be shown
in a similar way.

\section{Bipartite pure state disentanglement}

To gain some insight into the disentanglement process, the relatively simple
case of a bipartite system in a pure state $\left\vert \psi\right\rangle $ is
considered. Subsystems are labeled as 'a' and 'b'. With the help of the
Schmidt decomposition, $\left\vert \psi\right\rangle $ can be expressed as%
\begin{equation}
\left\vert \psi\right\rangle =%
{\displaystyle\sum\limits_{l=1}^{d_{\mathrm{m}}}}
q_{l}\left\vert l,l\right\rangle \;, \label{Schmidt decomposition BP}%
\end{equation}
where $d_{\mathrm{m}}=\min\left(  d_{\mathrm{a}},d_{\mathrm{b}}\right)  $, the
coefficients $q_{l}$ are non-negative real numbers, the tensor
product$\ \left\vert l\right\rangle _{\mathrm{a}}\otimes\left\vert
l\right\rangle _{\mathrm{b}}$ is denoted by $\left\vert l,l\right\rangle \ $,
and $\left\{  \left\vert l\right\rangle _{\mathrm{a}}\right\}  $ ($\left\{
\left\vert l\right\rangle _{\mathrm{b}}\right\}  $) is an orthonormal basis
spanning the Hilbert space $H_{\mathrm{a}}$\ ($H_{\mathrm{b}}$) of subsystem a
(b). The normalization condition reads $\left\langle \psi\right.  \left\vert
\psi\right\rangle =L_{2}=1$, where the $n$'th moment $L_{n}$ is defined by%
\begin{equation}
L_{n}=%
{\displaystyle\sum\limits_{l=1}^{d_{\mathrm{m}}}}
q_{l}^{n}\;. \label{L_n BP}%
\end{equation}

For a product state, for which $q_{l}=\delta_{l,l_{0}}$, where $l_{0}%
\in\left\{  1,2,\cdots,d_{\mathrm{m}}\right\}  $, $\tau_{\mathrm{ab}}$ obtains
its minimum value of $\tau_{\mathrm{ab}}=0$ [see Eqs. (\ref{C(O_a,O_b)}) and
(\ref{tau_12})]. The maximum value of $\tau_{\mathrm{ab}}$, which is given by
$\tau_{\mathrm{ab}}=\eta_{\mathrm{ab}}\left(  d_{\mathrm{m}}^{2}-1\right)
\left(  2/d_{\mathrm{m}}\right)  ^{2}$, is obtained when $q_{l}=d_{\mathrm{m}%
}^{-1/2}$ [maximum entropy state, see Eq. (\ref{tau_12})]. The constant
$\eta_{\mathrm{ab}}$ is chosen to be given by%
\begin{equation}
\eta_{\mathrm{ab}}=\frac{d_{\mathrm{m}}^{2}}{4\left(  d_{\mathrm{m}}%
^{2}-1\right)  }\;. \label{eta_ab}%
\end{equation}
For this choice $\tau_{\mathrm{ab}}$ is bounded between zero and unity.

Consider for simplicity the case where the Hamiltonian vanishes, i.e.
$\mathcal{H}=0$. For that case the modified Schr\"{o}dinger equation
(\ref{MSE}) for $d_{\mathrm{m}}\geq3$ yields%
\begin{equation}
\frac{\mathrm{d\log}q_{l}}{\mathrm{d}t}=4\gamma_{\mathrm{D}}\eta_{\mathrm{ab}%
}K_{l}^{\left(  3\right)  }\;, \label{d q_l / dt BP}%
\end{equation}
where the so-called capitalistic function $K_{l}^{\left(  m\right)  }$ is
given by $K_{l}^{\left(  m\right)  }=q_{l}^{2\left(  m-1\right)  }-L_{2m}$.
For $d_{\mathrm{m}}=2$ the factor $4$ in Eq. (\ref{d q_l / dt BP}) is replaced
by $12$. The identity $K_{l}^{\left(  m\right)  }=\partial H^{\left(
m\right)  }/\partial q_{l}$, where the potential function $H^{\left(
m\right)  }$ is given by%
\begin{equation}
H^{\left(  m\right)  }=\frac{1+m\left(  1-L_{2}\right)  }{2m}L_{2m}\;,
\label{H^(m)}%
\end{equation}
implies that $L_{2m}$ (i.e. $L_{6}$ for $m=3$) monotonically increases in time
(recall the normalization condition $L_{2}=1$). A similar derivation, based on
the operator $\mathcal{Q}^{\left(  \mathrm{S}\right)  }$, leads to a set of
equations of motion similar to (\ref{d q_l / dt BP}), but with $m=2$
\cite{Buks_2300103}. For that case $L_{4}$ monotonically increases in time
[see Eq. (\ref{H^(m)})].

\begin{figure}[ptb]
\begin{center}
\includegraphics[width=3.1in,keepaspectratio]{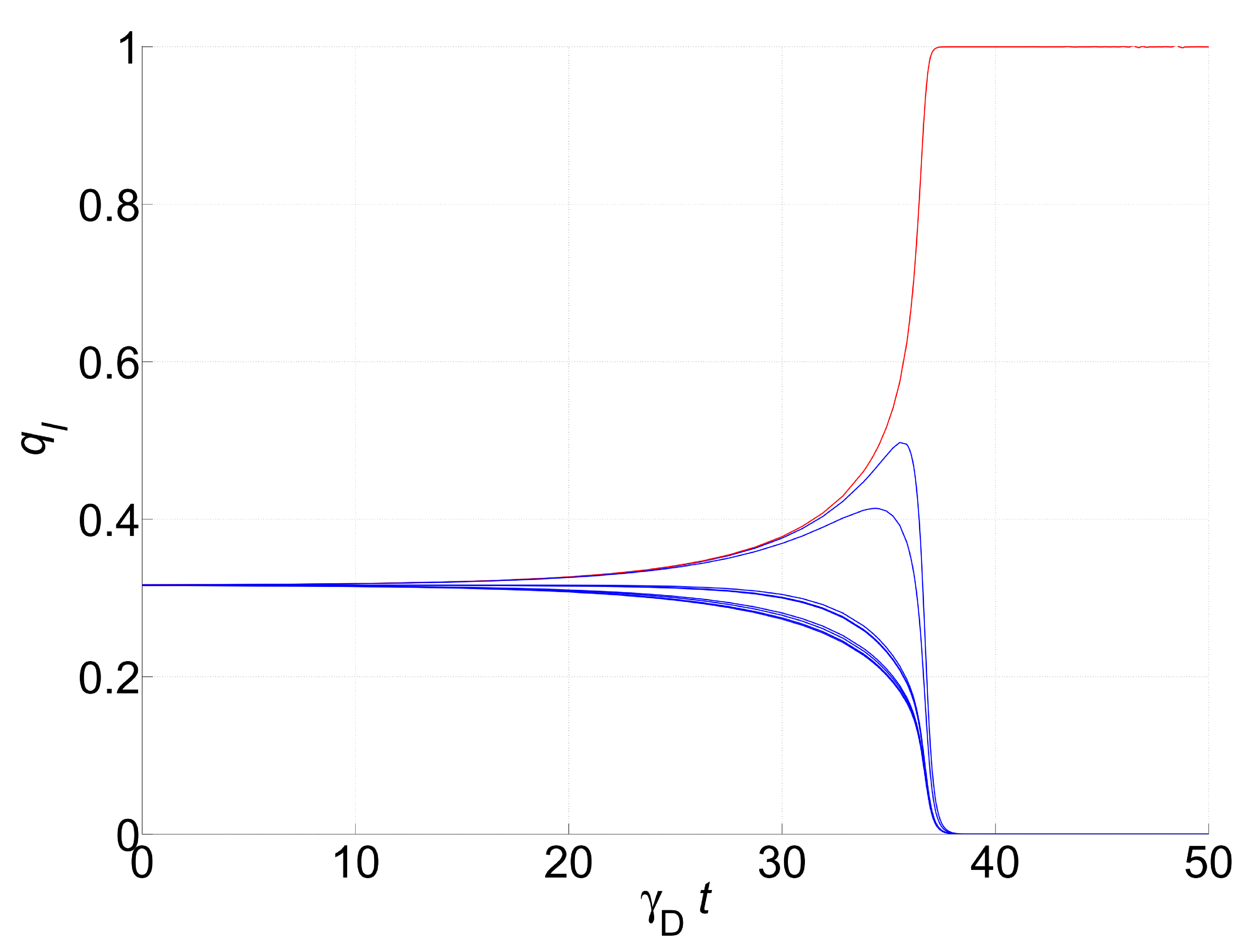}
\end{center}
\caption{{}Pure bipartite disentanglement. The time evolution of the
coefficients $q_{l}$ is calculated using Eq. (\ref{d q_l / dt BP}) for the
case $d_{\mathrm{m}}=10$ and $\gamma\eta_{\mathrm{ab}}=1$. The coefficients
$q_{l_{0}}$, which initially, at time $t=0$, is the largest one, i.e.
$q_{l_{0}}=\max\left\{  q_{l}\right\}  $, is represented by the red curve.}%
\label{Figdqdt}%
\end{figure}

The following holds [see Eq. (\ref{d q_l / dt BP})]%
\begin{equation}
\frac{\mathrm{d\log}\frac{q_{l^{\prime}}}{q_{l^{\prime\prime}}}}{\mathrm{d}%
t}=4\gamma_{\mathrm{D}}\eta_{\mathrm{ab}}\left(  q_{l^{\prime}}^{2\left(
m-1\right)  }-q_{l^{\prime\prime}}^{2\left(  m-1\right)  }\right)  \;.
\label{ql' / ql'' BP}%
\end{equation}
For both cases $\mathcal{Q}^{\left(  \mathrm{S}\right)  }$ (for which $m=2$)
and $\mathcal{Q}^{\left(  \mathrm{D}\right)  }$ (for which $m=3$), time
evolution governed by Eq. (\ref{ql' / ql'' BP}) gives rise to disentanglement.
Consider the case where initially, at time $t=0$, $q_{l_{0}}=\max\left\{
q_{l}\right\}  $ for a unique positive integer $l_{0}\in\left\{
1,2,\cdots,d_{\mathrm{m}}\right\}  $. As can be seen from Eq.
(\ref{ql' / ql'' BP}), for this case $\left\vert \psi\right\rangle $\ evolves
into the product state $\left\vert l_{0},l_{0}\right\rangle $ in the long time
limit, i.e. $q_{l}\rightarrow\delta_{l,l_{0}}$\ for $t\rightarrow\infty$. This
behavior is demonstrated by the plot shown in Fig. \ref{Figdqdt}.

\section{Two spin 1/2}

While thermalisation increases entropy, disentanglement decreases it (as is
demonstrated by the plot shown in Fig. \ref{Figdqdt}) \cite{Gharibyan_032125}.
The interplay between thermalisation and disentanglement is explored below
using a relatively simple system composed of two spins 1/2. For this case
$d_{\mathrm{a}}=d_{\mathrm{b}}=2$, and $\eta_{\mathrm{ab}}=1/3$ [see Eq.
(\ref{eta_ab})]. The angular momentum vector operator of spin $\mathrm{L}$ is
denoted by $\mathbf{S}_{\mathrm{L}}=\left(  S_{\mathrm{Lx}},S_{\mathrm{Ly}%
},S_{\mathrm{Lz}}\right)  $, where $\mathrm{L}\in\left\{  \mathrm{a}%
,\mathrm{b}\right\}  $.

Consider first the case where the system is in a pure state $\left\vert
\psi\right\rangle $ given by $\left\vert \psi\right\rangle =q_{00}\left\vert
00\right\rangle +q_{01}\left\vert 01\right\rangle +q_{10}\left\vert
10\right\rangle +q_{11}\left\vert 11\right\rangle $, where the ket vector
$\left\vert \sigma_{\mathrm{b}}\sigma_{\mathrm{a}}\right\rangle $ is an
eigenvector of $\left(  1/2\right)  \left(  1-\left(  2/\hbar\right)
S_{\mathrm{az}}\right)  $ and of $\left(  1/2\right)  \left(  1-\left(
2/\hbar\right)  S_{\mathrm{bz}}\right)  $, with eigenvalues $\sigma
_{\mathrm{a}}\in\left\{  0,1\right\}  $ and $\sigma_{\mathrm{b}}\in\left\{
0,1\right\}  $, respectively. For this case Eq. (\ref{tau_12}) yields
(hereafter it is assumed that disentanglement is generated by the operator
$\mathcal{Q}^{\left(  \mathrm{D}\right)  }$)%
\begin{equation}
\tau_{\mathrm{ab}}=\frac{8\left\vert \mathcal{D}\right\vert ^{2}\left(
1+2\left\vert \mathcal{D}\right\vert ^{2}\right)  }{3}\;,\label{tau TS}%
\end{equation}
where $\mathcal{D}=q_{00}q_{11}-q_{01}q_{10}$ (note that $\left\vert
D\right\vert ^{2}\leq1/4$ \cite{Wootters_2245}).

To explore the interplay between thermalisation and disentanglement, the
system's state is henceforth allowed to be mixed. Consider the case where the
Hamiltonian is given by $\mathcal{H}=-\hbar\omega_{\mathrm{B}}P_{\mathrm{B}}$,
where $\omega_{\mathrm{B}}$ is a positive constant, and $P_{\mathrm{B}%
}=\left\vert \psi_{\mathrm{B}}\right\rangle \left\langle \psi_{\mathrm{B}%
}\right\vert $ is a projection operator associated with the fully entangled
Bell singlet state $\left\vert \psi_{\mathrm{B}}\right\rangle =2^{-1/2}\left(
\left\vert 01\right\rangle -\left\vert 10\right\rangle \right)  $. As can be
verified using Eq. (\ref{MME}), the modified master equation for this case has
a fixed point given by%
\begin{equation}
\rho_{\mathrm{s}}=\frac{1+\kappa\left(  1-4P_{\mathrm{B}}\right)  }{4}\;,
\end{equation}
where the real variable $\kappa$ is found by solving
\begin{equation}
\log\frac{1-3\kappa}{1+\kappa}=\hbar\omega_{\mathrm{B}}\beta+\frac
{4\eta_{\mathrm{ab}}\gamma_{\mathrm{D}}\kappa}{\gamma_{\mathrm{H}}%
}\;.\label{eq kappa}%
\end{equation}

In the limit $\hbar\omega_{\mathrm{B}}\beta\gg1$ where thermalisation
dominates, Eq. (\ref{eq kappa}) yields $\kappa\simeq-1+4\exp\left(
-\hbar\omega_{\mathrm{B}}\beta+4\gamma_{\mathrm{D}}/\gamma_{\mathrm{H}%
}\right)  $ (i.e. $\rho_{\mathrm{s}}\simeq P_{\mathrm{B}}$). For this limit
the ground state $\left\vert \psi_{\mathrm{B}}\right\rangle $, which is fully
entangled, is nearly fully occupied, and the density matrix $\rho_{\mathrm{s}%
}$ represents a nearly pure state. In the opposite limit $\hbar\omega
_{\mathrm{B}}\beta\ll1$ where disentanglement dominates, Eq. (\ref{eq kappa})
yields $\kappa\simeq-\left(  1/4\right)  \hbar\omega_{\mathrm{B}}\beta\left(
1+\gamma_{\mathrm{D}}/\gamma_{\mathrm{H}}\right)  ^{-1}$ (i.e. $\rho
_{\mathrm{s}}\simeq1/4$). In this limit the density matrix $\rho_{\mathrm{s}}$
represents a nearly fully mixed and fully disentangled state.

\begin{figure}[ptb]
\begin{center}
\includegraphics[width=3.1in,keepaspectratio]{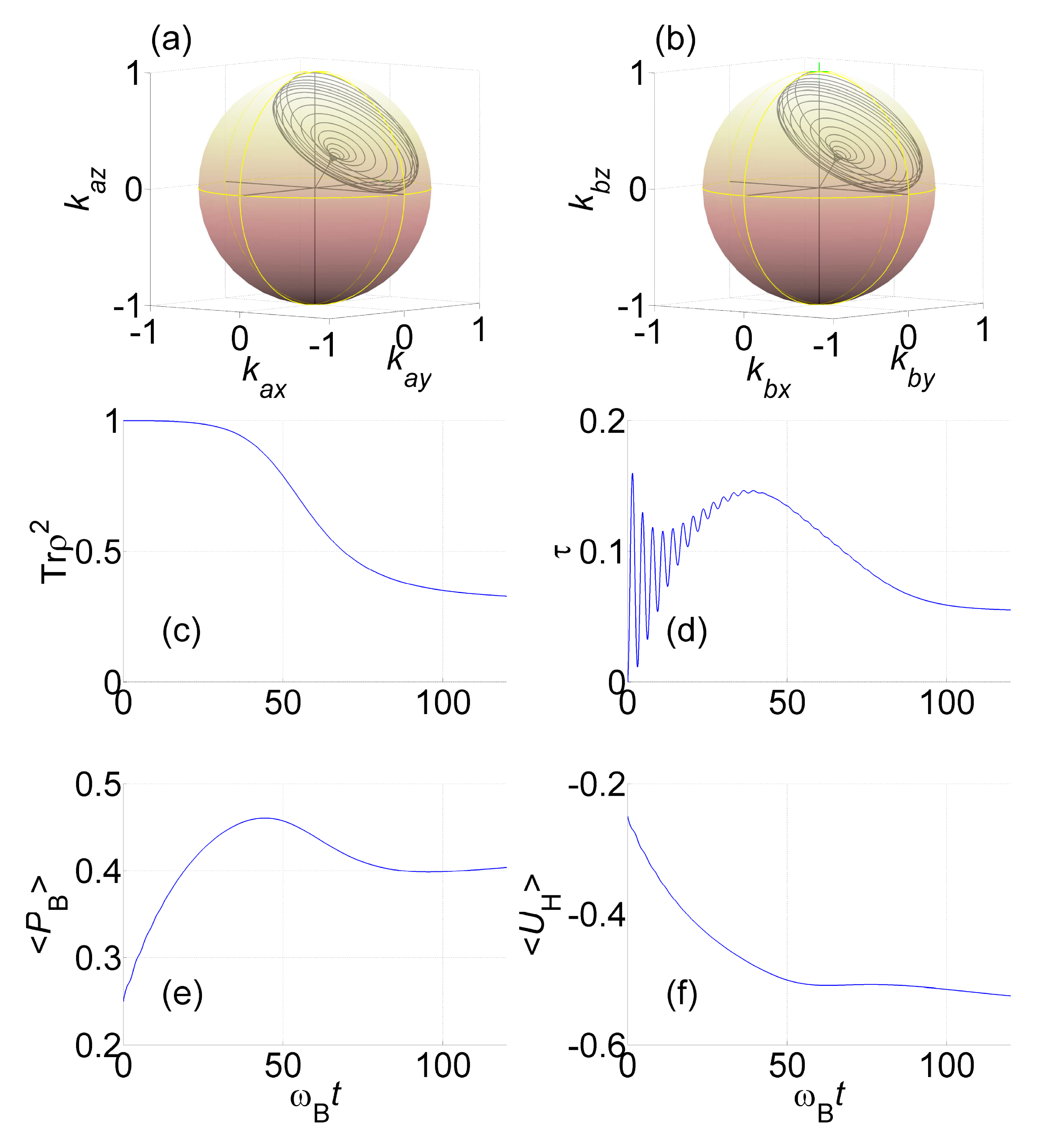}
\end{center}
\caption{{}Bell singlet state. The time evolution of the single spin Bloch
vectors $\mathbf{k}_{\mathrm{a}}$ and $\mathbf{k}_{\mathrm{b}}$ is shown in
(a) and (b), respectively. Initial values for $\mathbf{k}_{\mathrm{a}}$ and
$\mathbf{k}_{\mathrm{b}}$ are denoted by green cross symbols. (c) The purity
$\operatorname{Tr}\rho^{2}$. (d) The entanglement variable $\tau$. (e) The
expectation value $\left\langle P_{\mathrm{B}}\right\rangle $ of the
projection $P_{\mathrm{B}}=\left\vert \psi_{\mathrm{B}}\right\rangle
\left\langle \psi_{\mathrm{B}}\right\vert $. (f) The expectation value
$\left\langle \mathcal{U}_{\mathrm{H}}\right\rangle $\ of the Helmholtz free
energy. Assumed parameters' values are $\gamma_{\mathrm{H}}/\omega
_{\mathrm{B}}=0.005$, $\gamma_{\mathrm{D}}/\omega_{\mathrm{B}}=0.05$ and
$\hbar\omega_{\mathrm{B}}\beta=10$.}%
\label{FigBell}%
\end{figure}

An example for time evolution, which is obtained by numerically integrating
the modified master equation (\ref{MME}), is shown in Fig. \ref{FigBell}.
Assumed parameters' values are listed in the figure caption.

\section{Truncation approximation}

The simplest physical system suitable for the exploration of disentanglement
is the above-discussed two spin 1/2 system. For some cases, further
simplification can be achieved by implementing a truncation approximation.

\begin{figure}[ptb]
\begin{center}
\includegraphics[width=3.1in,keepaspectratio]{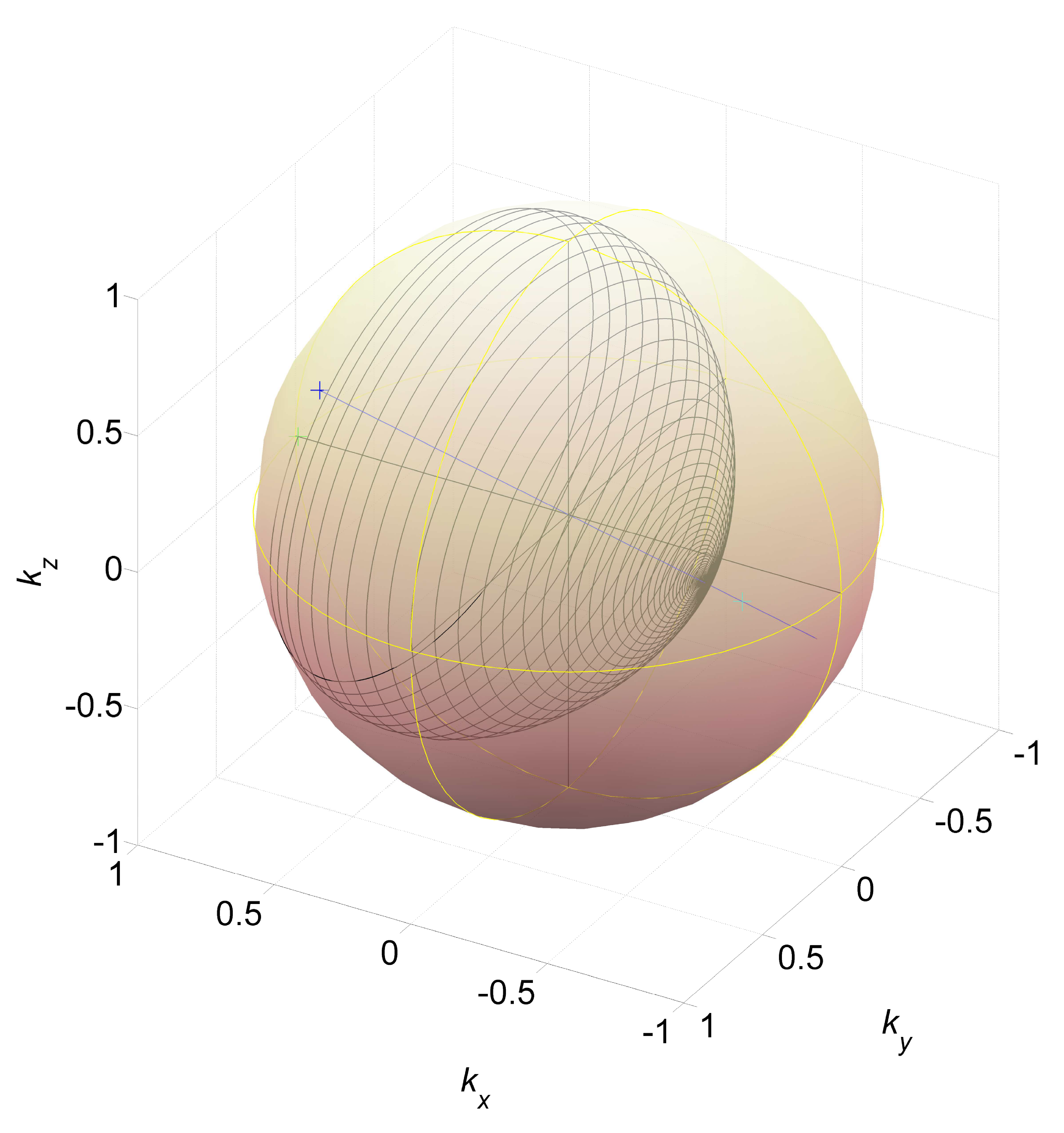}
\end{center}
\caption{{}Truncation approximation. The green cross symbol represents the
initial value of the Bloch vector $\mathbf{k}=\mu\mathbf{\hat{n}}$, the blue
line is parallel to $\boldsymbol{\omega}_{\mathrm{E}}$, the blue cross symbol
represents the unit vector ${\omega}_{\mathrm{E}}^{-1}
\boldsymbol{\omega}_{\mathrm{E}}$, and the cyan cross symbol represents the
point $-{\omega}_{\mathrm{E}}^{-1}\tanh\left(  \beta\hbar\omega_{\mathrm{E}
}/2\right)  \boldsymbol{\omega}_{\mathrm{E}}$ (steady state thermal
equilibrium in the absence of disentanglement). Assumed parameters' values are
$\gamma_{\mathrm{D}}/\gamma_{\mathrm{H}}=1$, $\boldsymbol{\omega}_{\mathrm{E}
}/\gamma_{\mathrm{H}}=10^{2}\left(  1,1,1\right)  $ and $\beta\hbar
\omega_{\mathrm{E}}=1$.}%
\label{FigTrunc}%
\end{figure}

Let $\mathcal{H}$ be the Hamiltonian of a two spin 1/2 system. The matrix
representation of $\mathcal{H}$ in the basis $\left\{  \left\vert
00\right\rangle ,\left\vert 01\right\rangle ,\left\vert 10\right\rangle
,\left\vert 11\right\rangle \right\}  $ is assumed to be given by%
\begin{equation}
\hbar^{-1}\mathcal{H}\dot{=}\Omega=\left(
\begin{array}
[c]{cccc}%
\frac{\omega_{\mathrm{s}}}{2} & 0 & 0 & 0\\
0 & \Omega_{22} & \Omega_{23} & 0\\
0 & \Omega_{32} & \Omega_{33} & 0\\
0 & 0 & 0 & \frac{\omega_{\mathrm{s}}}{2}%
\end{array}
\right)  \;,
\end{equation}
where the central $2\times2$ block of $\Omega$ is given by%
\begin{equation}
\left(
\begin{array}
[c]{cc}%
\Omega_{22} & \Omega_{23}\\
\Omega_{32} & \Omega_{33}%
\end{array}
\right)  =\frac{\boldsymbol{\omega}_{\mathrm{E}}\cdot\boldsymbol{\sigma}}%
{2}\;, \label{Omega CB}%
\end{equation}
both the scalar $\omega_{\mathrm{s}}$ and the vector $\boldsymbol{\omega
}_{\mathrm{E}}=\left(  \omega_{\mathrm{E}x},\omega_{\mathrm{E}y}%
,\omega_{\mathrm{E}z}\right)  $ are real, and $\boldsymbol{\sigma}=\left(
\sigma_{x},\sigma_{y},\sigma_{z}\right)  $ is the Pauli matrix vector.
Consider the case where $\hbar\omega_{\mathrm{s}}\beta\gg1$. For this case,
for which the population of the states $\left\vert 00\right\rangle $ and
$\left\vert 11\right\rangle $ is low, a truncation approximation can be
employed. In this approximation it is assumed that the system's Hilbert space
is spanned by the vector states $\left\vert 01\right\rangle $ and $\left\vert
10\right\rangle $. Note that in the truncation approximation, $\tau
_{\mathrm{ab}}$ for pure states is given by Eq. (\ref{tau TS}), with
$\mathcal{D}=q_{01}q_{10}$.

The truncated density matrix is expressed as%
\begin{equation}
\rho=\left(
\begin{array}
[c]{cccc}%
0 & 0 & 0 & 0\\
0 & \rho_{22} & \rho_{23} & 0\\
0 & \rho_{32} & \rho_{33} & 0\\
0 & 0 & 0 & 0
\end{array}
\right)  \;, \label{rho truncation}%
\end{equation}
where the central $2\times2$ block of $\rho$ is given by%
\begin{equation}
\left(
\begin{array}
[c]{cc}%
\rho_{22} & \rho_{23}\\
\rho_{32} & \rho_{33}%
\end{array}
\right)  =\frac{1+\mu\mathbf{\hat{n}}\cdot\boldsymbol{\sigma}}{2}\;,
\label{rho TST}%
\end{equation}
where $\mu\in\left[  0,1\right]  $ is real, and $\mathbf{\hat{n}}=\left(
n_{x},n_{y},n_{z}\right)  $ is a unit vector (i.e. $\mathbf{\hat{n}}%
\cdot\mathbf{\hat{n}}=1$). Note that $\operatorname{Tr}\rho^{2}=\left(
1/2\right)  \left(  1+\mu^{2}\right)  $. As can be seen from Eq.
(\ref{rho TST}), in the truncation approximation the system's state is
describe by a single real 3-dimensionla Bloch vector $\mathbf{k}%
=\mu\mathbf{\hat{n}}$.

With the help of the truncated density matrix $\rho$ (\ref{rho truncation}),
one finds that the matrix $\Theta^{\left(  \mathrm{D}\right)  }=\gamma
_{\mathrm{D}}\left(  \mathcal{Q}^{\left(  \mathrm{D}\right)  }\rho
+\rho\mathcal{Q}^{\left(  \mathrm{D}\right)  }-2\left\langle \mathcal{Q}%
^{\left(  \mathrm{D}\right)  }\right\rangle \rho\right)  $ is given by%
\begin{equation}
\Theta^{\left(  \mathrm{D}\right)  }=\left(
\begin{array}
[c]{cccc}%
0 & 0 & 0 & 0\\
0 & \Theta_{22}^{\left(  \mathrm{D}\right)  } & \Theta_{23}^{\left(
\mathrm{D}\right)  } & 0\\
0 & \Theta_{32}^{\left(  \mathrm{D}\right)  } & \Theta_{33}^{\left(
\mathrm{D}\right)  } & 0\\
0 & 0 & 0 & 0
\end{array}
\right)  \;,
\end{equation}
where the central $2\times2$ block of $\Theta^{\left(  \mathrm{D}\right)  }$
is given by%
\begin{equation}
\left(
\begin{array}
[c]{cc}%
\Theta_{22}^{\left(  \mathrm{D}\right)  } & \Theta_{23}^{\left(
\mathrm{D}\right)  }\\
\Theta_{32}^{\left(  \mathrm{D}\right)  } & \Theta_{33}^{\left(
\mathrm{D}\right)  }%
\end{array}
\right)  =\mathbf{k}^{\left(  \mathrm{D}\right)  }\cdot\boldsymbol{\sigma}\;,
\end{equation}
$\mathbf{k}^{\left(  \mathrm{D}\right)  }=-\left(  2/3\right)  \gamma
_{\mathrm{D}}\mu\left(  n_{x}\left(  N_{\perp}^{2}-1\right)  ,n_{y}\left(
N_{\perp}^{2}-1\right)  ,n_{z}N_{\perp}^{2}\right)  $, and $N_{\perp}^{2}
=\mu^{2}\left(  1-n_{z}^{2}\right)  $. The following holds%
\begin{equation}
\left\langle \mathcal{Q}^{\left(  \mathrm{D}\right)  }\right\rangle
=\frac{1+2\mu^{2}+\mu^{2}n_{z}^{2}\left(  \mu^{2}n_{z}^{2}-4\right)  }{3}\;,
\label{<Q^(D)>}%
\end{equation}
thus for a given $\mu$, the expectation value $\left\langle \mathcal{Q}%
^{\left(  \mathrm{D}\right)  }\right\rangle =\operatorname{Tr}\left(
\mathcal{Q}^{\left(  \mathrm{D}\right)  }\rho\right)  $\ is bounded by
$\left\langle \mathcal{Q}^{\left(  \mathrm{D}\right)  }\right\rangle
\in\left[  \left(  1/3\right)  \left(  1-\mu^{2}\right)  ^{2},\left(
1/3\right)  \left(  1+2\mu^{2}\right)  \right]  $. The expectation value
$\left\langle \mathcal{Q}^{\left(  \mathrm{D}\right)  }\right\rangle $ obtains
its minimum value $\left\langle \mathcal{Q}^{\left(  \mathrm{D}\right)
}\right\rangle =0$ for $n_{z}^{2}=1$ and $\mu=1$. These two points (north and
south poles of the Bloch sphere) represent fully disentangled states.

The entropy matrix $S=-\log\rho$ is given by%
\begin{equation}
S=\left(
\begin{array}
[c]{cccc}%
0 & 0 & 0 & 0\\
0 & s_{22} & s_{23} & 0\\
0 & s_{32} & s_{33} & 0\\
0 & 0 & 0 & 0
\end{array}
\right)  \;,
\end{equation}
where the central $2\times2$ block of $S$ is given by%
\begin{equation}
\left(
\begin{array}
[c]{cc}%
s_{22} & s_{23}\\
s_{32} & s_{33}%
\end{array}
\right)  =-\frac{1-\mathbf{\hat{n}}\cdot\boldsymbol{\sigma}}{2}\log\frac
{1-\mu}{2}-\frac{1+\mathbf{\hat{n}}\cdot\boldsymbol{\sigma}}{2}\log\frac
{1+\mu}{2}\;. \label{entropy operator}%
\end{equation}
Note that the expectation value $\left\langle S\right\rangle $, which is given
by%
\begin{equation}
\left\langle S\right\rangle =-\frac{1-\mu}{2}\log\frac{1-\mu}{2}-\frac{1+\mu
}{2}\log\frac{1+\mu}{2}\;,
\end{equation}
is bounded by $\left\langle S\right\rangle \in\left[  0,\log2\right]  $. With
the help of Eq. (\ref{entropy operator}) one finds that the central $2\times2$
block of $\left(  S\rho+\rho S-2\left\langle S\right\rangle \rho\right)  $ is
given by $\mathbf{k}^{\left(  \mathrm{S}\right)  }\cdot\boldsymbol{\sigma}$,
where $\mathbf{k}^{\left(  \mathrm{S}\right)  }=-\left(  1-\mu^{2}\right)
\left(  \tanh^{-1}\mu\right)  \mathbf{\hat{n}}$ [recall the identity
$\log\left(  \left(  1-\mu\right)  /\left(  1+\mu\right)  \right)
=-2\tanh^{-1}\mu$].

The central $2\times2$ block of $\left(  \mathcal{H}\rho+\rho\mathcal{H}%
-2\left\langle \mathcal{H}\right\rangle \rho\right)  $ is given by
$\mathbf{k}^{\left(  \mathrm{\Omega}\right)  }\cdot\boldsymbol{\sigma}$, where
$\mathbf{k}^{\left(  \mathrm{\Omega}\right)  }=\left(  \hbar/2\right)  \left(
\boldsymbol{\omega}_{\mathrm{E}}\mathbf{-}\mu^{2}\left(  \boldsymbol{\omega
}_{\mathrm{E}}\cdot\mathbf{\hat{n}}\right)  \mathbf{\hat{n}}\right)  $ [see
Eq. (\ref{Omega CB}), and recall the identity $\left(
\boldsymbol{\sigma}\cdot\mathbf{a}\right)  \left(  \boldsymbol{\sigma}\cdot
\mathbf{b}\right)  =\mathbf{a}\cdot\mathbf{b}+i\boldsymbol{\sigma}\cdot\left(
\mathbf{a}\times\mathbf{b}\right)  $], thus the central $2\times2$ block of
$\Theta^{\left(  \mathrm{H}\right)  }=\gamma_{\mathrm{H}}\left(
\mathcal{Q}^{\left(  \mathrm{H}\right)  }\rho+\rho\mathcal{Q}^{\left(
\mathrm{H}\right)  }-2\left\langle \mathcal{Q}^{\left(  \mathrm{H}\right)
}\right\rangle \rho\right)  $ is given by $\mathbf{k}^{\left(  \mathrm{H}%
\right)  }\cdot\boldsymbol{\sigma}$, where%
\begin{equation}
\frac{\mathbf{k}^{\left(  \mathrm{H}\right)  }}{\gamma_{\mathrm{H}}}%
=\frac{\beta\hbar\boldsymbol{\omega}_{\mathrm{E}}}{2}+\frac{2\left(  1-\mu
^{2}\right)  \tanh^{-1}\mu-\beta\hbar\mu^{2}\left(
\boldsymbol{\omega }_{\mathrm{E}}\cdot\mathbf{\hat{n}}\right)  }%
{2}\mathbf{\hat{n}}\;.
\end{equation}
The expectation value $\left\langle \Theta\right\rangle $ is given by
$\left\langle \Theta\right\rangle =\gamma_{\mathrm{H}}\left\langle
\mathcal{Q}^{\left(  \mathrm{H}\right)  }\right\rangle +\gamma_{\mathrm{D}%
}\left\langle \mathcal{Q}^{\left(  \mathrm{D}\right)  }\right\rangle $, where%
\begin{equation}
\left\langle \mathcal{Q}^{\left(  \mathrm{H}\right)  }\right\rangle =\frac
{\mu\beta\hbar\left(  \boldsymbol{\omega}_{\mathrm{E}}\cdot\mathbf{\hat{n}%
}\right)  }{2}+\frac{1-\mu}{2}\log\frac{1-\mu}{2}+\frac{1+\mu}{2}\log
\frac{1+\mu}{2}\;, \label{<Q_H>}%
\end{equation}
and $\left\langle \mathcal{Q}^{\left(  \mathrm{D}\right)  }\right\rangle $ is
given by Eq. (\ref{<Q^(D)>}). The expectation value $\left\langle
\mathcal{Q}^{\left(  \mathrm{H}\right)  }\right\rangle $ is minimized at
thermal equilibrium, for which $\boldsymbol{\omega}_{\mathrm{E}}%
\cdot\mathbf{\hat{n}}=-\omega_{\mathrm{E}}$ (i.e. $\mathbf{\hat{n}}$ is
anti-parallel to $\boldsymbol{\omega}_{\mathrm{E}}$) and $\mu=\tanh\left(
\beta\hbar\omega_{\mathrm{E}}/2\right)  $ [see Eq. (\ref{<Q_H>})].

The modified master equation (\ref{MME}) yields an equation of motion for the
3-dimensional Bloch vector $\mathbf{k}=\mu\mathbf{\hat{n}}$ given by%
\begin{equation}
\frac{\mathrm{d}\mathbf{k}}{\mathrm{d}t}=\boldsymbol{\omega}_{\mathrm{E}%
}\times\mathbf{k}-2\left(  \mathbf{k}^{\left(  \mathrm{H}\right)  }%
+\mathbf{k}^{\left(  \mathrm{D}\right)  }\right)  \;. \label{dk/dt}%
\end{equation}
An example of numerical integration of Eq. (\ref{dk/dt}) is shown in Fig.
\ref{FigTrunc}. This example demonstrates a disentanglement-induced shift of a
steady state fixed point away from thermal equilibrium (which is represented
by a cyan cross symbol). Assumed parameters are listed in the figure caption.

\section{Mean field approximation}

When disentanglement is sufficiently efficient (i.e. $\gamma_{\mathrm{D}}$ is
sufficiently large), the equations of motion generated by the modified master
equation (\ref{MME}) can be simplified by employing the mean field
approximation (MFA).

To demonstrate the MFA for the two spin 1/2 system, consider the case where
spin a has a relatively low angular Larmor frequency $\omega_{\mathrm{a}}$, in
comparison with the angular Larmor frequency $\omega_{\mathrm{b}}$ of spin b,
which is externally driven. For this case the Hamiltonian $\mathcal{H}$ of the
closed system is assumed to be given by%
\begin{equation}
\mathcal{H}=\omega_{\mathrm{a}}S_{\mathrm{az}}+\omega_{\mathrm{b}%
}S_{\mathrm{bz}}+\frac{\omega_{1}\left(  S_{\mathrm{b}+}+S_{\mathrm{b}%
-}\right)  }{2}+V\;,
\end{equation}
where the driving amplitude and angular frequency are denoted by $\omega_{1}$
and $\omega_{\mathrm{p}}=\omega_{\mathrm{b}}+\Delta$, respectively ($\Delta$
is the driving detuning), the operators $S_{\mathrm{a}\pm}$ are given by
$S_{\mathrm{a}\pm}=S_{\mathrm{ax}}\pm iS_{\mathrm{ay}}$, and the rotated
operators $S_{\mathrm{b}\pm}$ are given by $S_{\mathrm{b}\pm}=\left(
S_{\mathrm{bx}}\pm iS_{\mathrm{by}}\right)  e^{\pm i\omega_{\mathrm{p}}t}$.
The dipolar coupling term $V$ is given by $V=g\hbar^{-1}\left(  S_{\mathrm{a+}%
}+S_{\mathrm{a-}}\right)  S_{\mathrm{bz}}$, where $g$ is a coupling rate. The
largest effect of dipolar coupling occurs when the Hartmann--Hahn matching
condition $\omega_{\mathrm{a}}=\omega_{\mathrm{R}}$ is satisfied, where
$\omega_{\mathrm{R}}=\sqrt{\omega_{1}^{2}+\Delta^{2}}$ is the Rabi angular
frequency \cite{Hartmann1962,Yang_1,Levi_053516}.

Instead of employing the (nonlinear in $\rho$) operator $\mathcal{Q}^{\left(
\mathrm{H}\right)  }$, damping for this case is taken into account by adding
on the right hand side of the modified master equation (\ref{MME}) a Lindblad
superoperator $\mathcal{L}$ , which is linear in $\rho$, and which is given by
\cite{carmichael2009open}%
\begin{align}
\mathcal{L}  &  =\sum_{\mathrm{L}\in\left\{  \mathrm{a},\mathrm{b}\right\}
}\frac{\left(  \hat{n}_{0}^{\left(  \mathrm{L}\right)  }+1\right)  \Gamma
_{1}^{\left(  \mathrm{L}\right)  }}{4}\mathcal{D}_{\rho}\left(  \sigma
_{-}^{\left(  \mathrm{L}\right)  }\right)  +\frac{\hat{n}_{0}^{\left(
\mathrm{L}\right)  }\Gamma_{1}^{\left(  \mathrm{L}\right)  }}{4}%
\mathcal{D}_{\rho}\left(  \sigma_{+}^{\left(  \mathrm{L}\right)  }\right)
\nonumber\\
&  +\frac{\left(  2\hat{n}_{0}^{\left(  \mathrm{L}\right)  }+1\right)
\Gamma_{\varphi}^{\left(  \mathrm{L}\right)  }}{2}\mathcal{D}_{\rho}\left(
\sigma_{z}^{\left(  \mathrm{L}\right)  }\right)  \;,\nonumber\\
&  \label{superoperator}%
\end{align}
where the Lindbladian $\mathcal{D}_{\rho}\left(  X\right)  $ for an operator
$X$\ is given by%
\begin{equation}
\mathcal{D}_{\rho}\left(  X^{{}}\right)  =X^{{}}\rho X^{\dag}-\frac{X^{\dag
}X^{{}}\rho+\rho X^{\dag}X^{{}}}{2}\;, \label{Lindbladian}%
\end{equation}
the matrices $\sigma_{-}$ and $\sigma_{+}$ are given by $\sigma_{\pm}%
=\sigma_{x}\pm i\sigma_{y}$, and $\sigma_{x}$, $\sigma_{y}$ and $\sigma_{z}$
are Pauli matrices. The positive damping rates $\Gamma_{1}^{\left(
\mathrm{L}\right)  }$ and $\Gamma_{\varphi}^{\left(  \mathrm{L}\right)  }$,
and the thermal occupation factor $\hat{n}_{0}^{\left(  \mathrm{L}\right)  }$,
are related to the longitudinal $T_{1}^{\left(  \mathrm{L}\right)  }$ and the
transverse $T_{2}^{\left(  \mathrm{L}\right)  }$ relaxation times, and to the
thermal equilibrium spin polarization $k_{z0}^{\left(  \mathrm{L}\right)  }$,
by $1/T_{1}^{\left(  \mathrm{L}\right)  }=\Gamma_{1}^{\left(  \mathrm{L}%
\right)  }\left(  2\hat{n}_{0}^{\left(  \mathrm{L}\right)  }+1\right)  $,
$1/T_{2}^{\left(  \mathrm{L}\right)  }=\left(  \Gamma_{1}^{\left(
\mathrm{L}\right)  }/2+\Gamma_{\varphi}^{\left(  \mathrm{L}\right)  }\right)
\left(  2\hat{n}_{0}^{\left(  \mathrm{L}\right)  }+1\right)  $ and
$-1/k_{z0}^{\left(  \mathrm{L}\right)  }=2\hat{n}_{0}^{\left(  \mathrm{L}%
\right)  }+1$.


Equations of motion for the single spin Bloch vectors $\mathbf{k}_{\mathrm{a}%
}$ and $\mathbf{k}_{\mathrm{b}}$ are derived from the modified master equation
(\ref{MME}) using Eq. (\ref{eom G}) of appendix \ref{AppHSF}. In the MFA Eq.
(\ref{MFA}) of appendix \ref{AppHSF} yields%
\begin{equation}
\frac{\mathrm{d}\mathbf{k}_{\mathrm{a}}}{\mathrm{d}t}=\left(
\begin{array}
[c]{c}%
-\omega_{\mathrm{a}}k_{\mathrm{ay}}-\frac{k_{\mathrm{ax}}}{T_{2\mathrm{a}}}\\
\omega_{\mathrm{a}}k_{\mathrm{ax}}-gk_{\mathrm{az}}k_{\mathrm{bz}}%
-\frac{k_{\mathrm{ay}}}{T_{2\mathrm{a}}}\\
gk_{\mathrm{ay}}k_{\mathrm{bz}}-\frac{k_{\mathrm{az}}-k_{\mathrm{z0a}}%
}{T_{1\mathrm{a}}}%
\end{array}
\right)  \;, \label{dka/dt}%
\end{equation}
and%
\begin{equation}
\frac{\mathrm{d}\mathbf{k}_{\mathrm{b}}}{\mathrm{d}t}=\left(
\begin{array}
[c]{c}%
-\Delta k_{\mathrm{by}}-gk_{\mathrm{ax}}k_{\mathrm{by}}-\frac{k_{\mathrm{bx}}%
}{T_{2\mathrm{b}}}\\
\Delta k_{\mathrm{bx}}-\omega_{1}k_{\mathrm{bz}}+gk_{\mathrm{ax}%
}k_{\mathrm{bx}}-\frac{k_{\mathrm{by}}}{T_{2\mathrm{b}}}\\
\omega_{1}k_{\mathrm{by}}-\frac{k_{\mathrm{bz}}-k_{\mathrm{z0b}}%
}{T_{1\mathrm{b}}}%
\end{array}
\right)  \;. \label{dkb/dt}%
\end{equation}

\begin{figure}[ptb]
\begin{center}
\includegraphics[width=3.1in,keepaspectratio]{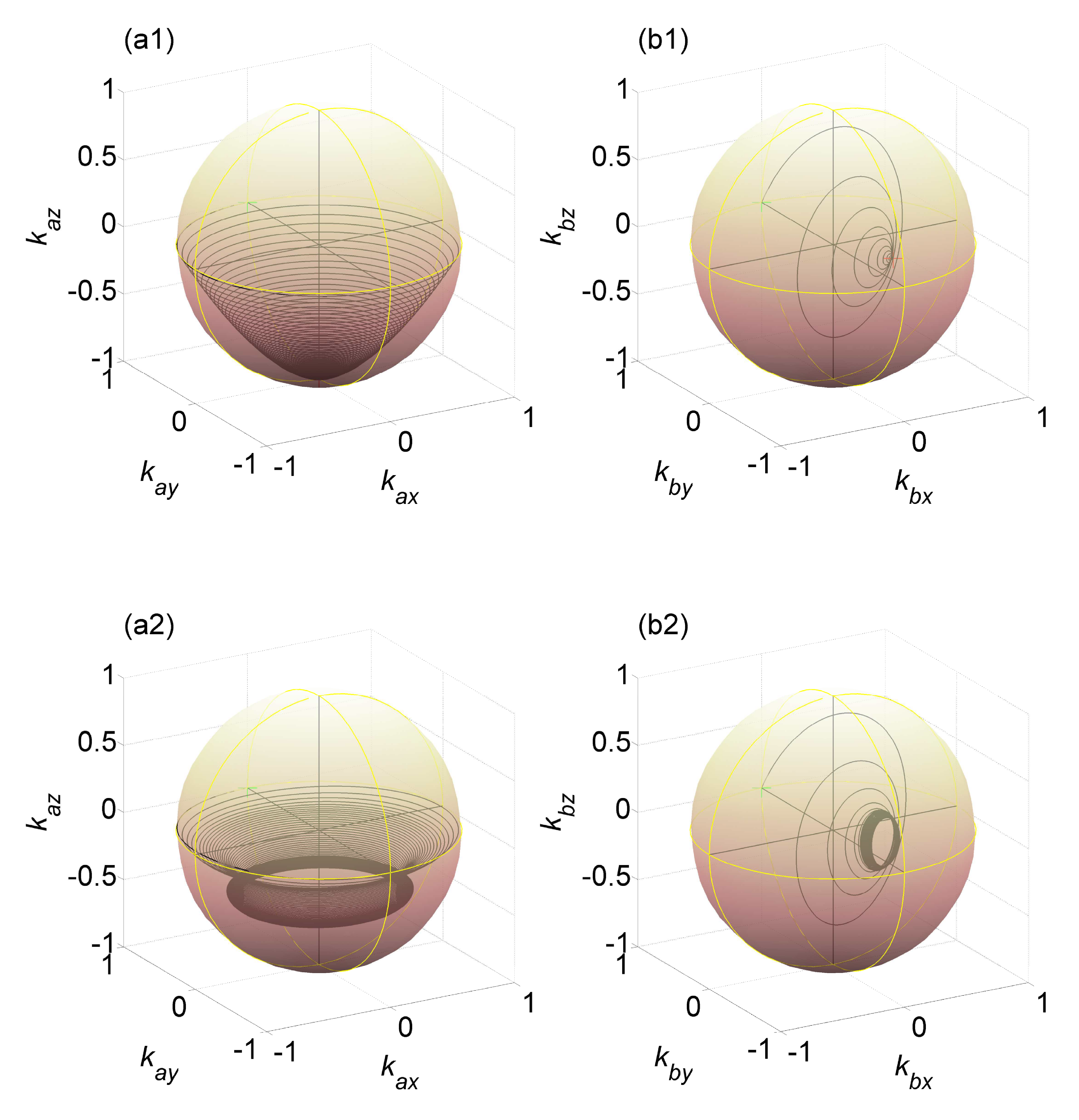}
\end{center}
\caption{Mean field approximation. The time evolution of the single spin Bloch
vector $\mathbf{k}_{\mathrm{a}}$ ($\mathbf{k}_{\mathrm{b}}$) is shown in the
plots labeled by the letter a (b). For plots (a1) and (b1) the spins are
decoupled (i.e. $g=0$), whereas $g/\omega_{\mathrm{a}}=0.1$ for plots (a2) and
(b2). Other assumed parameters' values are $\Delta/\omega_{\mathrm{a}}
=\sin\left(  \pi/8\right)  $, $\omega_{1}/\omega_{\mathrm{a}}=\cos\left(
\pi/8\right)  $ (note that the Hartmann--Hahn matching condition is
satisfied), $\Gamma_{1}^{\left(  \mathrm{a}\right)  }/\omega_{\mathrm{a}
}=10^{-2}$, $\Gamma_{\varphi}^{\left(  \mathrm{a}\right)  }/\Gamma
_{1}^{\left(  \mathrm{a}\right)  }=10^{-1}$, $\Gamma_{1}^{\left(
\mathrm{b}\right)  }/\Gamma_{1}^{\left(  \mathrm{a}\right)  }=10$,
$\Gamma_{\varphi}^{\left(  \mathrm{b}\right)  }/\Gamma_{\varphi}^{\left(
\mathrm{a}\right)  }=10$, $\hat{n}_{0}^{\left(  \mathrm{a}\right)  }=0.005$
and $\hat{n}_{0}^{\left(  \mathrm{b}\right)  }=0.0001$. Initial values for
$\mathbf{k}_{\mathrm{a}}$ and $\mathbf{k}_{\mathrm{b}}$ are denoted by green
cross symbols.}%
\label{FigMFA}%
\end{figure}

The plots shown in Fig. \ref{FigMFA} exhibit the time evolution of the single
spin Bloch vectors $\mathbf{k}_{\mathrm{a}}$ and $\mathbf{k}_{\mathrm{b}}$,
where $\mathbf{k}_{\mathrm{L}}=\left(  \left(  1/2\right)  \left\langle
S_{\mathrm{L}+}+S_{\mathrm{L}-}\right\rangle ,\left(  -i/2\right)
\left\langle S_{\mathrm{L}+}-S_{\mathrm{L}-}\right\rangle ,\left\langle
S_{\mathrm{Lz}}\right\rangle \right)  $, and where $\mathrm{L}\in\left\{
\mathrm{a},\mathrm{b}\right\}  $. The case $g=0$ (i.e. no dipolar coupling) is
represented by the plots (a1) and (b1). For this case the steady state is a
fixed point, which is labeled by a red cross symbol in Fig. \ref{FigMFA}(a1)
and (b1). On the other hand, in the presence of sufficiently strong dipolar
coupling, the steady state becomes a limit cycle, as is demonstrated by the
plots shown in Fig. \ref{FigMFA}(a2) and (b2), for which $g/\omega
_{\mathbf{a}}=0.1$. The limit cycle angular frequency is close to
$\omega_{\mathrm{a}}$.

\section{Discussion and summary}

The above-discussed MFA greatly simplifies the analysis. For the two spin 1/2
system, the modified master equation (\ref{MME}) leads to a set [see Eq.
(\ref{eom G})] of $4^{2}-1=15$ real equations of motion (generalized Bloch
equation) for $15$ real variables (generalized Bloch vector). On the other
hand, for the same system, the MFA leads to a set of $2^{2}-1+2^{2}-1=6$ real
equations of motion for $6$ real variables [see Eqs. (\ref{dka/dt}) and
(\ref{dkb/dt})].

As is demonstrated by the plots shown in Fig. \ref{FigMFA}, the nonlinear
terms in Eqs. (\ref{dka/dt}) and (\ref{dkb/dt}) can give rise to a limit cycle
steady state. Such limit cycle steady states cannot be obtained from the GKSL
master equation, which is linear in $\rho$ (see appendix B of Ref.
\cite{Levi_053516}). On the other hand, in spite of the linear dependency of
the GKSL equation on $\rho$, some theoretical studies have revealed nonlinear
dynamics that is derived from this GKSL master equation
\cite{breuer2002theory,Drossel_217,Hicke_024401,Klobus_034201,Hush_061401}.
However, the origin of such nonlinearity is the assumption that entanglement
between subsystems can be disregarded. It has remained unclear how such an
assumption can be justified in the framework of standard QM. On the other
hand, this assumption represents a limiting case for the modified master
equation (\ref{MME}), for which disentanglement is sufficiently efficient.

The limit cycle steady state shown in Fig. \ref{FigMFA}(a2) and (b2) can occur
only when the driving detuning $\Delta$ is positive (i.e. driving is
blue-detuned). This behavior demonstrates that disentanglement can give rise
to detuning asymmetry. On the other hand, in the absence of disentanglement,
the system's response is theoretically expected to be an even function of the
detuning $\Delta$ [e.g. see Eq. (4) of Ref. \cite{Attrash_054001}]. Many
examples can be found in the published literature for a profound detuning
asymmetry observed in spin systems under nutation driving or dynamical
decoupling. In most papers the presented asymmetry is not discussed, however,
a paper from 1955 \cite{Redfield_1787}, and another one from 2005
\cite{Fedaruk_473}, explicitly state that the observed asymmetry is
theoretically unexpected. Moreover, limit cycle steady states are
experimentally observed in systems of correlated spins \cite{Anderson_1788}.
Further study is needed to explore possible connections between experimentally
observed nonlinear dynamics in spin systems \cite{Desvaux_50} and disentanglement.

In summary, the spontaneous disentanglement hypothesis is inherently
falsifiable, because it yields predictions, which are experimentally
distinguishable from predictions obtained from standard QM. In particular, as
was discussed above, the experimental observation of a limit cycle steady
state in a system having a Hilbert space of finite dimensionality (i.e. a spin
system) may provide a supporting evidence for the spontaneous disentanglement
hypothesis. Moreover, such an experimental observation may yield some insight
related to the question 'what determines the disentanglement rate
$\gamma_{\mathrm{D}}$?', which has remained entirely open.

\appendix

\section{Hilbert space factorization}

\label{AppHSF}

Consider a $d_{\mathrm{H}}$-dimensional Hilbert space, where $d_{\mathrm{H}%
}\in\left\{  2,3,\cdots\right\}  $ is finite. The generalized Gell-Mann set
$G=\left\{  \lambda_{l}\right\}  $, which spans the SU($d_{\mathrm{H}}$) Lie
algebra, contains $d_{\mathrm{H}}^{2}-1$ square $d_{\mathrm{H}}\times
d_{\mathrm{H}}$ Hermitian matrices. For the case $d_{\mathrm{H}}=2$
($d_{\mathrm{H}}=3$) the 3 (8) set elements are called Pauli (Gell-Mann)
matrices. The Generalized Gell-Mann matrices are traceless, i.e.
$\operatorname{Tr}\lambda_{l}=0$, and they satisfy the orthogonality relation%
\begin{equation}
\frac{\operatorname{Tr}\left(  \lambda_{l^{\prime}}\lambda_{l^{\prime\prime}%
}\right)  }{2}=\delta_{l^{\prime},l^{\prime\prime}}\;. \label{GM OR}%
\end{equation}

Unless $d_{\mathrm{H}}$ is prime, it can be factored as $d_{\mathrm{H}%
}=d_{\mathrm{a}}d_{\mathrm{b}}$, where $d_{\mathrm{a}}>1$ and $d_{\mathrm{b}%
}>1$ are both integers. The two subsystems corresponding to the factorization
\cite{Carroll_022213} are labelled as 'a' and 'b', respectively. The
generalized Gell-Mann $d_{\mathrm{L}}\times d_{\mathrm{L}}$ matrices
corresponding to subsystem $\mathrm{L}$, where $\mathrm{L}\in\left\{
\mathrm{a},\mathrm{b}\right\}  $, are denoted by $\lambda_{l}^{\left(
\mathrm{L}\right)  }$, where $l\in\left\{  1,2,\cdots,d_{\mathrm{L}}%
^{2}-1\right\}  $. For a given factorization, consider the set of
$d_{\mathrm{H}}^{2}-1$ matrices $G^{\left(  \mathrm{ab}\right)  }=\left\{
\Gamma_{a}^{\left(  \mathrm{a}\right)  }\otimes\Gamma_{b}^{\left(
\mathrm{b}\right)  }\right\}  -\left\{  \Gamma_{0}^{\left(  \mathrm{a}\right)
}\otimes\Gamma_{0}^{\left(  \mathrm{b}\right)  }\right\}  $, where
$a\in\left\{  0,1,2,\cdots,d_{\mathrm{a}}^{2}-1\right\}  $ and $b\in\left\{
0,1,2,\cdots,d_{\mathrm{b}}^{2}-1\right\}  $. For subsystem $\mathrm{L}$,
where $\mathrm{L}\in\left\{  \mathrm{a},\mathrm{b}\right\}  $, the matrix
$\Gamma_{0}^{\left(  \mathrm{L}\right)  }$ is defined by $\Gamma_{0}^{\left(
\mathrm{L}\right)  }=\left(  2^{1/4}/d_{\mathrm{L}}^{1/2}\right)
I_{\mathrm{L}}$,\ where $I_{\mathrm{L}}$ is the $d_{\mathrm{L}}\times
d_{\mathrm{L}}$ identity matrix, and for $l\in\left\{  1,2,\cdots
,d_{\mathrm{L}}^{2}-1\right\}  $\ the matrix $\Gamma_{l}^{\left(
\mathrm{L}\right)  }$ is defined by $\Gamma_{l}^{\left(  \mathrm{L}\right)
}=2^{-1/4}\lambda_{l}^{\left(  \mathrm{L}\right)  }$.

With the help of the Kronecker matrix product identities $\operatorname{Tr}%
\left(  A\otimes B\right)  =\operatorname{Tr}A\operatorname{Tr}B$ and $\left(
A\otimes B\right)  \left(  C\otimes D\right)  =\left(  AC\right)
\otimes\left(  BD\right)  $, one finds that the set $G^{\left(  \mathrm{ab}%
\right)  }$ shares two properties with the Gell-Mann set $G$ of the
$d_{\mathrm{H}}$-dimensional Hilbert space. The first one is tracelessness
$\operatorname{Tr}G_{a,b}=0$ for any $G_{a,b}\equiv\Gamma_{a}^{\left(
\mathrm{a}\right)  }\otimes\Gamma_{b}^{\left(  \mathrm{b}\right)  }\in
G^{\left(  \mathrm{ab}\right)  }$ [recall that $G_{0,0}\notin G^{\left(
\mathrm{ab}\right)  }$], and the second one is orthogonality [see Eq.
(\ref{GM OR})]%
\begin{equation}
\frac{\operatorname{Tr}\left(  G_{a^{\prime},b^{\prime}}G_{a^{\prime\prime
},b^{\prime\prime}}\right)  }{2}=\delta_{a^{\prime},a^{\prime\prime}}%
\delta_{b^{\prime},b^{\prime\prime}}\;. \label{G OR}%
\end{equation}

The set $G^{\left(  \mathrm{ab}\right)  }$ can be used to expand the entire
system density matrix $\rho$ (which is assume to be normalized, i.e.
$\operatorname{Tr}\rho=1$) as%
\begin{equation}
\rho=\sum_{a=0}^{d_{\mathrm{a}}^{2}-1}\sum_{b=0}^{d_{\mathrm{b}}^{2}-1}%
\frac{\left\langle G_{a,b}\right\rangle G_{a,b}}{2}\;, \label{rho G}%
\end{equation}
where $\left\langle O\right\rangle =\operatorname{Tr}\left(  O\rho\right)  $
for a given observable $O$. Partial trace is used to derive the reduced
density matrices $\rho_{\mathrm{a}}=\operatorname{Tr}_{\mathrm{b}}%
\rho=2^{-1/2}\sum_{a=0}^{d_{\mathrm{a}}^{2}-1}\left\langle \Gamma_{a}^{\left(
\mathrm{a}\right)  }\otimes I_{\mathrm{b}}\right\rangle \Gamma_{a}^{\left(
\mathrm{a}\right)  }$ and $\rho_{\mathrm{b}}=\operatorname{Tr}_{\mathrm{a}%
}\rho=2^{-1/2}\sum_{b=0}^{d_{\mathrm{b}}^{2}-1}\left\langle I_{\mathrm{a}%
}\otimes\Gamma_{b}^{\left(  \mathrm{b}\right)  }\right\rangle \Gamma
_{b}^{\left(  \mathrm{b}\right)  }$ [recall the identities $\operatorname{Tr}%
_{\mathrm{A}}\left(  A\otimes B\right)  =\operatorname{Tr}\left(  A\right)  B$
and $\operatorname{Tr}_{\mathrm{B}}\left(  A\otimes B\right)
=\operatorname{Tr}\left(  B\right)  A$]. Level of entanglement can be
characterized using the matrix $D=\rho-\rho_{\mathrm{a}}\otimes\rho
_{\mathrm{b}}$, which is given by%
\begin{align}
D  &  =\sum_{a=0}^{d_{\mathrm{a}}^{2}-1}\sum_{b=0}^{d_{\mathrm{b}}^{2}-1}%
\frac{\left(  \left\langle G_{a,b}\right\rangle -\left\langle \Gamma
_{a}^{\left(  \mathrm{a}\right)  }\otimes I_{\mathrm{b}}\right\rangle
\left\langle I_{\mathrm{a}}\otimes\Gamma_{b}^{\left(  \mathrm{b}\right)
}\right\rangle \right)  G_{a,b}}{2}\nonumber\\
&  =\sum_{a=1}^{d_{\mathrm{a}}^{2}-1}\sum_{b=1}^{d_{\mathrm{b}}^{2}-1}%
\frac{\left(  \left\langle \lambda_{a,b}\right\rangle -\left\langle
\lambda_{a}^{\left(  \mathrm{a}\right)  }\otimes I_{\mathrm{b}}\right\rangle
\left\langle I_{\mathrm{a}}\otimes\lambda_{b}^{\left(  \mathrm{b}\right)
}\right\rangle \right)  \lambda_{a,b}}{4}\;,\nonumber\\
&  \label{D=}%
\end{align}
where $\lambda_{a,b}=\lambda_{a}^{\left(  \mathrm{a}\right)  }\otimes
\lambda_{b}^{\left(  \mathrm{b}\right)  }$. For any product state $D=0$.
Alternatively, as can be seen from Eq. (\ref{D=}), the density matrix $\rho$
represents a product state if and only if $\operatorname{rank}\left\langle
G_{a,b}\right\rangle =1$ [this is proved by showing that the assumption
$\operatorname{rank}\left\langle G_{a,b}\right\rangle =1$, which implies that
the $d_{\mathrm{a}}^{2}\times d_{\mathrm{b}}^{2}$ matrix $\left\langle
G_{a,b}\right\rangle $ equals an outer product, leads to $\left\langle
\lambda_{a}^{\left(  \mathrm{a}\right)  }\otimes\lambda_{b}^{\left(
\mathrm{b}\right)  }\right\rangle /\left(  \left\langle \lambda_{a}^{\left(
\mathrm{a}\right)  }\otimes I_{\mathrm{b}}\right\rangle \left\langle
I_{\mathrm{a}}\otimes\lambda_{b}^{\left(  \mathrm{b}\right)  }\right\rangle
\right)  =1$].

\begin{figure}[ptb]
\begin{center}
\includegraphics[width=3.1in,keepaspectratio]{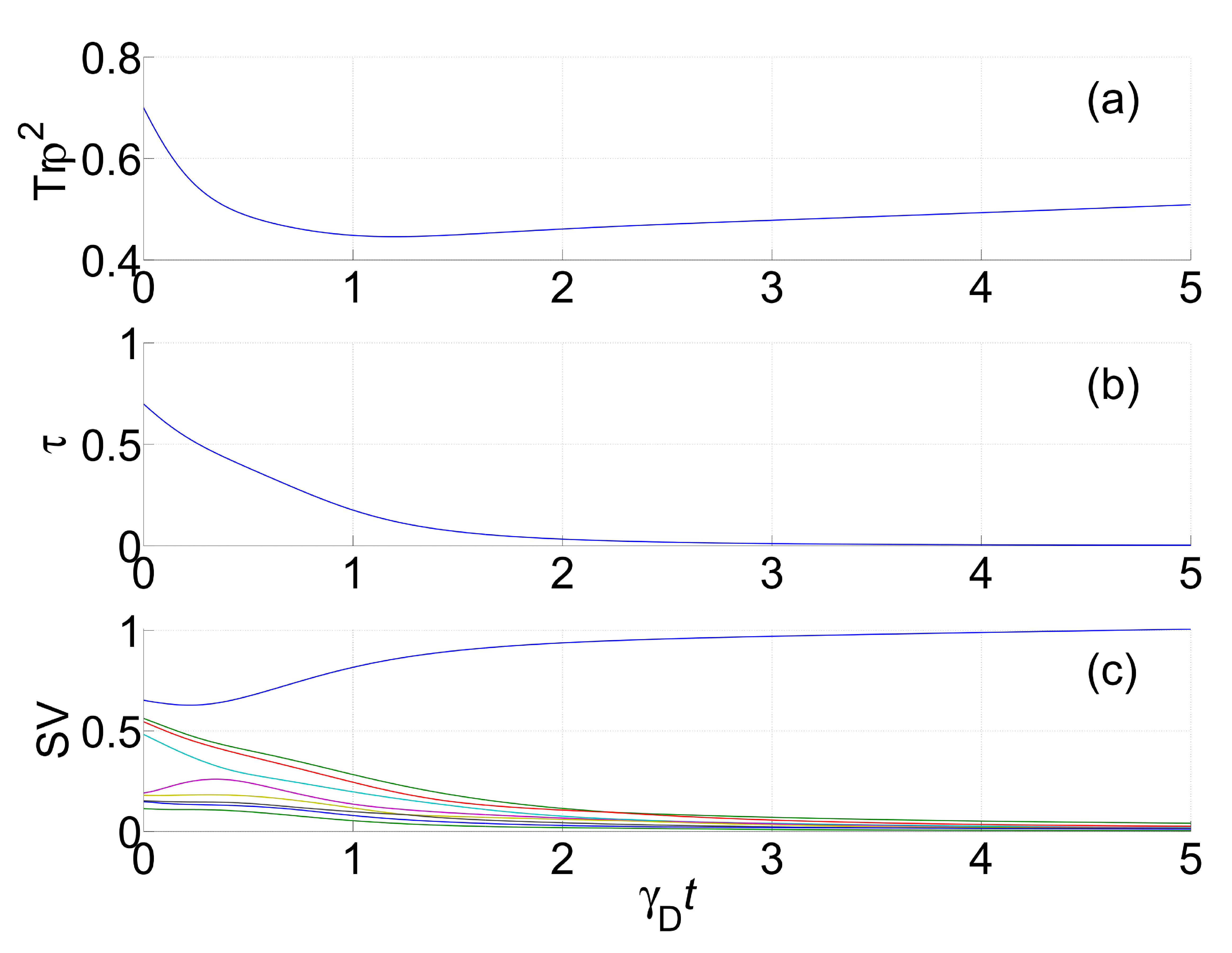}
\end{center}
\caption{{}The Bloch matrix $\left\langle G_{a,b}\right\rangle $. For this
example, $\mathcal{H}=0$, $\gamma_{\mathrm{H}}=0$, $d_{\mathrm{a}}=3$ and
$d_{\mathrm{b}}=4$. (a) Purity $\operatorname{Tr}\rho^{2}$. (b) Entanglement
$\tau$. (c) SV of the Bloch matrix $\left\langle G_{a,b}\right\rangle $.}%
\label{FigSVD}%
\end{figure}

The $d_{\mathrm{a}}^{2}\times d_{\mathrm{b}}^{2}$ matrix $\left\langle
G_{a,b}\right\rangle $ is henceforth referred to as the Bloch matrix. By
expanding the master equation as [see Eq. (\ref{G OR})]%
\begin{equation}
\frac{\mathrm{d}\rho}{\mathrm{d}t}=\sum_{a=0}^{d_{\mathrm{a}}^{2}-1}\sum
_{b=0}^{d_{\mathrm{b}}^{2}-1}\frac{\operatorname{Tr}\left(  \frac
{\mathrm{d}\rho}{\mathrm{d}t}G_{a,b}\right)  G_{a,b}}{2}\;, \label{MEG}%
\end{equation}
one finds that the time evolution of the Bloch matrix is governed by [see Eq.
(\ref{rho G})]%
\begin{equation}
\frac{\mathrm{d}\left\langle G_{a,b}\right\rangle }{\mathrm{d}t}%
=\operatorname{Tr}\left(  \frac{\mathrm{d}\rho}{\mathrm{d}t}G_{a,b}\right)
\;. \label{eom G}%
\end{equation}
In the mean field approximation (MFA) it is assumed that%
\begin{equation}
\left\langle G_{a,b}\right\rangle =\left\langle G_{a,0}\right\rangle
\left\langle G_{0,b}\right\rangle \;. \label{MFA}%
\end{equation}

To demonstrate the effect of spontaneous disentanglement on the time evolution
of the Bloch matrix $\left\langle G_{a,b}\right\rangle $, the case where the
Hamiltonian $\mathcal{H}$ vanishes, and $\gamma_{\mathrm{H}}=0$ (i.e. no
thermalisation) is considered. For the plots shown in Fig. \ref{FigSVD},
$d_{\mathrm{a}}=3$ and $d_{\mathrm{b}}=4$. The time evolution of (a) the
purity $\operatorname{Tr}\rho^{2}$, (b) the entanglement $\tau$, and (c) the
singular values (SV) of the Bloch matrix $\left\langle G_{a,b}\right\rangle $,
is evaluated by numerically integrating the master equation (\ref{MME}). Note
that, in the long time limit $\gamma_{\mathrm{D}}t\rightarrow\infty$, for
which entanglement if fully suppressed, i.e. $\tau=0$, the Bloch matrix
$\left\langle G_{a,b}\right\rangle $ has a single non-zero SV (i.e. its rank
becomes unity).

\bibliographystyle{ieeepes}
\bibliography{acompat,Eyal_Bib}

\end{document}